# Prediction of ferroelectricity-driven Berry curvature enabling charge- and spin-controllable photocurrent in tin telluride monolayers

Jeongwoo Kim[1,2†], Kyoung-Whan Kim[3,4†], Dongbin Shin[1], Sang-Hoon Lee[5], Jairo Sinova[4,6], Noejung Park[1] & Hosub Jin[1*]

[1]*Department of Physics, Ulsan National Institute of Science and Technology, Ulsan 44919, Korea*

[2]*Department of Physics, Incheon National University, Incheon 22012, Korea*

[3]*Center for Spintronics, Korea Institute of Science and Technology, Seoul 02792, Korea*

[4]*Institute of Physics, Johannes Gutenberg University Mainz, Mainz, 55099, Germany*

[5]*Korea Institute for Advanced Study, Seoul 02455, Korea*

[6]*Institute of Physics, Academy of Sciences of the Czech Republic, Cukrovarnická 10, 162 53 Praha 6, Czech Republic*

## Abstract

**In symmetry-broken crystalline solids, pole structures of Berry curvature (BC) can emerge, and they have been utilized as a versatile tool for controlling transport properties. For example, the monopole component of the BC is induced by the time-reversal symmetry breaking, and the BC dipole arises from a lack of inversion symmetry, leading to the anomalous Hall and nonlinear Hall effects, respectively. Based on first-principles calculations, we show that the ferroelectricity in a tin telluride monolayer produces a unique BC distribution, which offers charge- and spin-controllable photocurrents. Even with the sizable band gap, the ferroelectrically driven BC dipole is comparable to those of small-gap topological materials. By manipulating the photon handedness and the ferroelectric polarization, charge and spin circular photogalvanic currents are generated in a controllable manner. The ferroelectricity in group-IV monochalcogenide monolayers can be a useful tool to control the BC dipole and the nonlinear optoelectronic responses.**

[†]: *J. Kim and K.-W. Kim equally contributed to this work.*

**Introduction**

The concept of Berry curvature (BC) is becoming increasingly pertinent due to its central role in various topological phases and unusual transport phenomena.[1-5] Under symmetry-broken environments in crystalline solids, BC can emerge from the quantum geometry embedded in the electronic structure. It provides an effective magnetic field in momentum space and deforms the electron motion in real space, which becomes a primary origin of exotic transport properties, including various Hall effects.[6-13] As a representative example, when time-reversal symmetry is broken, the anomalous Hall effect arises from the net flux of the BC, which is known as the Berry phase.[4,12,14,15] Recently, the dipole component of the BC that can be induced by inversion asymmetry has been attracting increasing attention due to its potential for optoelectronic applications.[16-19] Under out-of-equilibrium electron distributions, the BC dipole can enable nonlinear optoelectronic transport, which has been realized in photogalvanic experiments.[20,21]

Since level crossing can generate a singular BC distribution, topological materials that possess small inverted band gaps or band crossings have been mostly studied as efficient platforms for hosting a large BC dipole.[20-26] For instance, a small-gap quantum spin Hall $WTe_2$ monolayer shows a large inter-band BC and its dipole is manipulated by an external electric field, resulting in the circular photogalvanic effect.[20] Tilted Weyl semimetals and pressurized BiTeI that is driven towards the topological phase transition regime also exhibit a large enhancement in the intra-band BC dipole, leading to the nonlinear Hall effect by generating a transverse photocurrent under linearly polarized light.[24,25] Despite the large BC dipoles in these topological materials, the prominent nonlinear optical properties are available only in response to low-frequency fields due to the small size of the band gap and the sub-band energy splitting. For high-frequency applications, the system requires a larger band gap corresponding to a higher photon energy. However, realizing a large BC dipole is challenging in large-band-gap systems because a large band gap impedes singular band inversion and crossing. Therefore, it is desirable to identify a new mechanism for producing a large amount of BC and its corresponding dipole, even in the presence of a relatively large band gap.

For BC engineering in large-band-gap systems, we suggest that ferroelectricity can serve as a tool for manipulating the BC distribution by providing an inversion-breaking order parameter. Using first-principles density functional theory (DFT), we demonstrate that the in-plane ferroelectricity in a SnTe monolayer exhibits a large BC distribution with a band gap of ~ 1 eV, which corresponds to the near-infrared or visible light range. Due to the ferroelectricity, a pair of positive and negative BC peaks is formed, naturally inducing a BC dipole. Microscopically, the ferroelectric displacement develops nearest-neighbour inter-orbital hopping channels that are otherwise forbidden by the structural symmetry and it efficiently mixes the orbital characteristics of the conduction and valence bands. Consequently, a pair of the opposite orbital angular momentum textures appears and is referred to as the orbital Rashba effect, to which the large BC dipole is primarily ascribed. In addition to the conventional application of the BC dipole for the nonlinear optoelectronics, we present an intriguing approach for controlling the spin and charge photocurrents either separately or simultaneously via ferroelectric polarization in cooperation with photon helicity. Considering the non-volatile switching of the electric polarization, the large BC and its dipole in large-gapped ferroelectric systems provide a new approach for multifunctional nonlinear optoelectronic and optospintronic applications.

**Results**

**Atomic and electronic structure of the SnTe monolayer.** The SnTe monolayer has a *Pmn2₁* space group due to the in-plane ferroelectricity.[27] It can be considered as a binary version

4of the phosphorene puckered structure, where Sn and Te atoms undergo opposite displacements along the [100] direction (the *x*-axis in Fig. 1a).[28-30] As a result, the SnTe monolayer has an in-plane ferroelectricity of 12.4 µC cm$^{-2}$ (Supplementary Fig. 1). A mirror symmetry ($M_{xz}$) and a glide mirror symmetry (*G*) exist, as illustrated in Fig. 1a. When the ferroelectric polarization is aligned along the *x*-axis, the electronic structure of the SnTe monolayer exhibits two valleys near points X and Y,[31] which are hereafter referred to as the X and Y valleys, respectively (Fig. 1b). The conduction and valence bands near the Fermi level (-1 to 1 eV) mostly originate from Sn- and Te-5*p* orbitals, respectively (Supplementary Fig. 2a). The lowest conduction bands (the highest valence bands) are derived from the Sn(Te)-$p_x$ orbital at the X valley, while they are derived from the $p_y$ orbital at the Y valley (Fig. 1b). Along the circular path from the X valley to the Y valley, the atomic orbitals are aligned along the radial direction for the lowest conduction band (~ 0.7 eV) and along the tangential direction for the second-lowest band (~ 1.1 eV) (for details, see Supplementary Fig. 2b-d).

The orbital splitting between the radial and tangential orbitals is described by a simple model Hamiltonian, which, in the following sections, shall play a crucial role in ferroelectrically driven BC dipoles. Conduction electrons near the X valley can be modelled by

$$H_0(\mathbf{k}) = E_X(\mathbf{k}) - J \cos 2\theta_\mathbf{k} \, \tau_z - J \sin 2\theta_\mathbf{k} \, \tau_x, \qquad (1)$$

where $\tau_x = |p_x\rangle\langle p_y| + |p_y\rangle\langle p_x|$, $\tau_y = -i|p_x\rangle\langle p_y| + i|p_y\rangle\langle p_x|$, and $\tau_z = |p_x\rangle\langle p_x| - |p_y\rangle\langle p_y|$ are the pseudospin Pauli matrices in the $p_{x/y}$-orbital basis, $E_X(\mathbf{k})$ is the kinetic energy in the form of $\hbar^2(\mathbf{k} - \mathbf{k}_X)^2/2m$ near the X valley, and $\mathbf{k}_X$ is the position of the X valley. The second and third terms in $H_0(\mathbf{k})$ models the momentum-dependent orbital splitting, where $2J > 0$ is the orbital splitting energy, and $\theta_\mathbf{k} = \arg(k_x + ik_y)$ makes the orbital structure consistent with the DFT calculations (Supplementary Figs. 2c and 2d).

Due to the orthorhombic structure induced by the in-plane ferroelectricity, the X and Y valleys are inequivalent to each other. For the X valley, unidirectional Rashba-type spin splitting is observed along the $k_y$-direction (Fig. 1c), which is generated by the combination of $x$-axis ferroelectric polarization and spin-orbit coupling (SOC).[32-34] The spin states are degenerate along the Γ-X line due to the mirror symmetry $M_{xz}$. In addition, for the spin expectation value at each spin split-off band, the glide mirror symmetry $G$ allows the out-of-plane component only, as shown in Fig. 1c.

**Ferroelectrically driven Berry curvature.** When the ferroelectric polarization breaks the inversion symmetry of the system, a unique polarization-dependent BC distribution can emerge. To elucidate the relation between the ferroelectricity and the BC in the SnTe monolayer, we calculate the BC distribution, namely, $\mathbf{\Omega}(\mathbf{k}) = \Omega(\mathbf{k})\mathbf{z}$, over the first Brillouin zone (see Methods). As shown in Fig. 2a, large BC peaks are established at the X valley. Interestingly, a pair of positive and negative BC peaks appears at the X valley, naturally linking to the BC dipole, which is shown later. Moreover, the overall sign changes by the polarization reversal (Supplementary Fig. 3); hence, direct coupling occurs between the BC and the ferroelectricity. Therefore, the ferroelectricity of the SnTe monolayer provides an efficient approach for producing a large BC distribution in a controllable manner.

With varying the magnitude of the ferroelectric polarization ($P$) or the SOC strength ($\lambda$), the BC distribution is investigated in Fig. 2b-e. The BCs are presented along the $k_y$-direction across the X valley (the blue dashed line in Fig. 2a). As $P$ increases, two neighbouring BC peaks with opposite signs develop and they gradually move away from each other (Fig. 2b). Once the ferroelectricity flips, the sign of the BC distribution also changes (Figs. 2b and 2c). In contrast, the SOC alters the BC insignificantly (Figs. 2d and 2e). The BC profile is well maintained regardless of the value of $\lambda$. The dependence on $P$ and $\lambda$ demonstrates that the BC is governed by the ferroelectricity rather than the SOC.

Based on the following symmetry argument, we can qualitatively interpret the behaviour of the BC distributions with respect to ferroelectricity, which is summarized in Figs. 2f and 2g. We decompose the BC for spin $\sigma(=\pm)$ bands in terms of $\Omega_\sigma(\mathbf{k})$ as $\Omega(\mathbf{k}) = \Omega_+(\mathbf{k}) + \Omega_-(\mathbf{k})$. Under the time-reversal ($\mathcal{T}$) and mirror-reflection ($M_{xz}$) operations, $\Omega_\sigma(\mathbf{k})$ is transformed as follows:

$$\Omega_\pm(\mathbf{k}) \xrightarrow{\mathcal{T}} -\Omega_\mp(-\mathbf{k}), \quad (2)$$

$$\Omega_\pm(k_x, k_y) \xrightarrow{M_{xz}} -\Omega_\mp(k_x, -k_y). \quad (3)$$

Eq. (3) follows from both the BC and the spin lying on the $xz$ mirror plane. As a result, $\Omega(\mathbf{k}) = -\Omega(-\mathbf{k})$ and $\Omega(k_x, k_y) = -\Omega(k_x, -k_y)$ due to the time-reversal symmetry and the mirror symmetry of the system, respectively. In the absence of $P$, the system recovers the inversion symmetry that leads to $\Omega_\pm(\mathbf{k}) = \Omega_\pm(-\mathbf{k})$. Consequently, $\Omega(\mathbf{k})$ is zero without $P$ (Fig. 2f). If we consider the ferroelectric reversal, $\Omega_\pm^{+P}(\mathbf{k}) = \Omega_\pm^{-P}(-\mathbf{k})$ because the $+P$ and $-P$ configurations are inversion partners. Combining this with Eq. (2) yields $\Omega^{+P}(\mathbf{k}) = -\Omega^{-P}(\mathbf{k})$ (Fig. 2g). The symmetry argument well explains the DFT calculation results that reveal the ferroelectrically coupled BC.

**Origin of the ferroelectricity-coupled Berry curvature.** By extending the Hamiltonian of Eq. (1), we now construct an analytic model that provides microscopic origin on how the BC distribution can couple with the ferroelectricity. The ferroelectric polarization introduces new hopping channels that, when combined with the orbital spitting in Eq. (1), are the primary origin of the BC arising in the SnTe monolayer.

Figures 3a and 3b describe the new hopping channels driven by the ferroelectric displacement; an effective inter-orbital hopping channel from the Sn-$p_x$ orbital to the nearest-neighbour Sn-$p_y$ orbital along the *y*-axis is activated by the in-plane ferroelectricity. When the ferroelectric polarization has stabilized, the asymmetric hopping integrals are developed in Fig. 3a. After integrating out the Te atoms, an effective anti-symmetric inter-orbital hopping occurs with a hopping amplitude of $t_{xy}^{\text{eff}}$ that is proportional to the ferroelectric polarization. Upon the reversal of the ferroelectric polarization, the sign of the effective hopping is reversed (Fig. 3b). Such a ferroelectrically driven hopping Hamiltonian for Sn can be expressed as $H_{\text{FE}}(\mathbf{k}) = it_{xy}^{\text{eff}}(|p_y\rangle\langle p_x| - |p_x\rangle\langle p_y|)\sin k_y a' \approx t_{xy}^{\text{eff}} a' k_y \tau_y$ near the X valley, where $a' = \sqrt{2}a$ is the distance between neighbouring Sn atoms (see Supplementary Note 1 for a more rigorous tight-binding approach and its justification by comparison with DFT calculations). A similar Hamiltonian can be derived for electrons in Te atoms as well.

The ferroelectrically driven model Hamiltonian is equivalent to

$$H_{\text{FE}}(\mathbf{k}) = \alpha_L k_y L_z = |\alpha_L|\mathbf{L}\cdot(\widehat{\mathbf{P}}\times\mathbf{k}), \tag{4}$$

since the orbital angular momentum operator, $L_z$, is represented by $\hbar\tau_y$ in the $p_{x/y}$-orbital subspace. Here, $\alpha_L = t_{xy}^{\text{eff}} a'/\hbar$ is proportional to the ferroelectric polarization. We assumed that the direction of the ferroelectric polarization ($\widehat{\mathbf{P}}$) is taken as $+\mathbf{x}$ without loss of generality. Eq. (4) is identical to the Rashba-type Hamiltonian[32,33] if we replace the spin angular momentum **S** by the orbital angular momentum **L**; therefore, it can be referred to as the orbital Rashba effect.[35] Here, we explicitly demonstrate the emergence of the orbital Rashba effect from the ferroelectrically allowed inter-orbital hybridizations in Fig. 3a, b. In the absence of the ferroelectric polarization, such inter-orbital hopping channels are cancelled out due to the inversion symmetry (see Supplementary Fig. 4). Consistently with the DFT calculations of Figs. 3c and 3d, the orbital Rashba effect produces the *z*-component orbital angular momentum texture that is an odd function of both $k_y$ and ferroelectric polarization. From the close correlation between BC and orbital angular momentum, as exemplified in graphene[8] and MoS$_2$[9], one can expect that the orbital Rashba effect can bring a similar BC distribution.

To examine the relation between the ferroelectricity and the BC, we diagonalize the total Hamiltonian $H_0(\mathbf{k}) + H_{\text{FE}}(\mathbf{k})$ and obtain the following expression for the BC:

$$\Omega(\mathbf{k}) = \frac{2\alpha_L J^2}{J_\mathbf{k}^3 \hbar} \partial_{k_x} \theta_\mathbf{k}, \tag{5}$$

where $J_\mathbf{k} = \sqrt{J^2 + (\alpha_L \hbar k_y)^2}$ is the modified orbital splitting (for details, see Methods and Supplementary Note 2). The linear $\alpha_L$-dependence of the BC implies that $\Omega(\mathbf{k})$ is induced by the ferroelectric polarization and it is switchable by reversing $P$. Since $\theta_\mathbf{k} = \arg(k_x + ik_y)$, the BC is an odd function of $k_y$, and thus it forms a dipole along the $y$ direction. This result indicates that the ferroelectrically driven orbital Rashba effect plays a central role in developing the BC dipole in the SnTe monolayer.

**Intra/inter-band Berry curvature dipoles and nonlinear responses.** The steep slope between the two adjacent and opposite BC peaks at the X valley gives rise to a large BC dipole, which induces nonlinear optoelectronic responses such as the nonlinear Hall effect and the circular photogalvanic effect.[16-18] By using DFT method, we calculate the intra-band BC ($\Omega_v$), inter-band BC ($\Omega_{vc}$) distributions, and their dipoles as shown in Fig. 4 (for the formalisms, see Methods). Due to the mirror symmetry $M_{xz}$, both the intra- and inter-band BC dipoles are composed of the $y$-component only. $\Omega_v$ and $\Omega_{vc}$ show similar distributions along the $k_y$-direction, leading to large BC dipoles. For a more accurate estimation, we performed our calculations with two exchange-correlation potentials, namely, the Perdew-Burke-Ernzerhof (PBE)[36] and Heyd-Scuseria-Ernzerho (HSE)[37] functionals, which yield the same qualitative trend.

When the SnTe monolayer is doped, the intra-band BC dipole $\mathbf{D}^{\text{intra}}(\mu)$ emerges and produces the nonlinear Hall current in response to a low-frequency photon.[18] By assuming that the chemical potential $\mu$ is close to the band edge $\mu_0$ of the parabolic band, we obtain $D_y^{\text{intra}}(\mu) = -\alpha_L(m/\pi J\hbar^3 k_X^2)|\mu - \mu_0|$ from Eq. (5). This induces a nonlinear Hall effect in the SnTe monolayer; a DC Hall current flows along the *x*-direction in response to the normal incident light with *y*-polarization and reveres its direction upon ferroelectric reversal.

For the pristine insulating phase, the inter-band BC dipole $\mathbf{D}^{\text{inter}}(\omega)$ determines the inter-band circular photogalvanic effect.[20] Provided that the incident photon energy $\hbar\omega$ is comparable to the direct bandgap $E_{\text{gap}}$ at the X valley, optoelectronic responses are mainly governed by the band-edge transitions. Then, our analytic formalism gives the following expression for the inter-band BC dipole:

$$D_y^{\text{inter}}(\omega) \propto \alpha_L \left(1 - \frac{E_{\text{gap}}}{\hbar\omega}\right), \tag{6}$$

whose complete expression is presented in Supplementary Eq. (20). The inset of Fig. 4d presents the frequency dependence of Eq. (6) which accords with our DFT results. The circular photogalvanic current is then calculated from the relation[20] $J_{y,\pm} = \pm(2\pi e^3 \tau E_0^2/\hbar^2)D_y^{\text{inter}}(\omega)$, where $\pm$ refers to the incident photon helicity, $\tau$ is the momentum relaxation time, and $E_0$ is the field amplitude of the light.

The order of magnitude of the BC dipoles in the SnTe monolayer (~ 0.1 Å) is comparable to that of the small-gap or gapless topological materials.[20,21] For instance, 0.1 Å of the inter-band BC dipole has been obtained in the WTe$_2$ monolayer under an external field of $E_z$ ~ 1.5 V nm$^{-1}$, giving rise to ~ 200 nA W$^{-1}$ circular photogalvanic current.[20] Therefore, the SnTe monolayer is a promising platform for exploring the BC-related nonlinear optoelectronic responses over a wide frequency range.

**Discussion**

When combined with the Rashba spin splitting that is caused by the ferroelectricity and the SOC, the unique BC structure yields additional fascinating optoelectronic responses in the SnTe monolayer; one can control the spin polarization of the photocurrent as well as its charge degree of freedom. According to the spin-resolved BC profiles that are depicted in Figs. 4a and 4b, the positive and negative BC peaks are dominated by the spin-up and spin-down components, respectively. Such strong coupling of the spin polarization and the BC distribution leads to the spin- and momentum-asymmetric circular dichroism in Fig. 5a. Via the combination of the circular dichroism and the large Rashba spin splitting, each spin split-off band can be selectively excited by circularly polarized light with normal incidence,[38] thereby producing a current-carrying non-equilibrium electron distribution (Fig. 5b). As a result, we can generate spin-polarized circular photogalvanic currents in the SnTe monolayer. Furthermore, such charge and spin photocurrents can be separately configured via circular dichroism and ferroelectric polarization.

Using the time-dependent DFT method to describe the non-equilibrium electron dynamics, we directly demonstrate the generation of charge and spin circular photogalvanic currents along with their possibility to be manipulated. When the SnTe monolayer is exposed to a time-varying circularly polarized electric field whose frequency is tuned to the band gap, the charge and spin currents are generated as shown in Fig. 5c. Although the time-dependent DFT method captures all opto-electronic/spintronic responses from first to higher order contributions, we identify non-zero direct current (DC) components by plotting guidelines in Fig. 5c, which are related to the BC dipole. We confirmed that the DC component of the current is consistent with the second order response theory[39] (see Supplementary Fig. 5). These charge and spin currents flow perpendicular to the ferroelectric polarization and vary by switching the photon helicity and ferroelectric polarization.

When the photon helicity changes along with the fixed ferroelectric polarization, the charge current flows backwards while the spin current is unaltered, i.e., the difference between the spin currents under the right- and left-handed circularly polarized lights is zero (Fig. 5d). Upon varying the photon helicity, the spin and the group velocity of the excited carrier are simultaneously reversed, thus affecting the charge current only (see the parabolic bands with black dots in Fig. 5g). In Fig. 5e, the charge and spin current differences have finite values as the ferroelectric polarization is flipped while the photon helicity is fixed; hence, both the charge and spin currents reverse their directions. Under the ferroelectric reversal, both the spin and the BC are flipped in the electronic structure. The crystal momentum of the excited carrier is then reversed while its spin direction is fixed, resulting in the reversal of both the charge and spin currents. If we change the ferroelectric polarization and the photon helicity at the same time, the spin current is reversed, whereas the charge current remains unaffected (Fig. 5f). Therefore, the charge and spin degrees of freedom in circular photogalvanic current can be readily controlled simultaneously and/or separately by means of the photon handedness and the ferroelectricity (Fig. 5g).

The switchable behaviour of the charge and spin currents depicted in Fig. 5g can also be understood by the following symmetry argument. The two oppositely polarized configurations (+$P$ and –$P$) are connected by the inversion operation (vertical white arrows). And the charge and spin currents are odd under the spatial inversion. Therefore, both charge and spin photocurrents should be reversed under the ferroelectric reversal. On the other hand, the time-reversal operation transforms the SnTe monolayer exposed to the right circularly-polarized light into the one to the left circularly-polarized light, and vice versa. Under the time reversal (horizontal white arrows), the charge current is odd while the spin current is even. As a result, the reversal of the photon helicity inverts the direction of the charge current only, remaining the spin current invariant. In addition to the time-dependent DFT calculations and the symmetry argument, we develop an analytic formalism for the charge and spin photogalvanic currents to verify their ferroelectric origin and the manoeuvrability by the photon handedness and the ferroelectricity [see Supplementary Eqs. (21), (22), (24), and (25)].

There are two side remarks. First, our main results are likely to be valid for thicker SnTe films in which the in-plane ferroelectricity has been reported to be retained.[27] Moreover, the series of group-IV monochalcogenide monolayers share the same ferroelectrically driven BC dipole features due to their similar electronic structures.[31] Second, the ferroelectrically driven BC dipole in the SnTe monolayer is different from the BC dipole proposed in the surface of the well-known topological crystalline insulator SnTe[18] where the singular BC distribution originates from the gapped and tilted surface Dirac cones. In our work, a sizeable BC dipole appears in a trivial and large insulating gap of the SnTe monolayer with the help of ferroelectricity.

In summary, we identified the fundamental relation between the ferroelectricity and the BC dipole, which is a counterpart of the well-known ferromagnetism and BC monopole coupling. Based on this finding, we demonstrated the possibility of generating and manipulating photocurrents via the ferroelectrically driven BC dipole in the SnTe monolayer. Despite the large gap in the SnTe monolayer, its intra- and inter-band BC dipoles were predicted to reach substantial values that are comparable to those of the experimentally measured $WTe_2$ monolayer. The anti-symmetric inter-orbital hopping that is induced by the ferroelectricity gives rise to the orbital Rashba effect, which plays an essential role in the BC dipole structure. In addition, we presented the charge and spin circular photogalvanic currents and, on the basis of the light handedness and ferroelectric polarization switching, we proposed a pragmatic scheme for simultaneously or independently controlling them. Through the large ferroelectrically driven and, thus, ferroelectrically controlled BC dipole, the SnTe monolayer can serve as a unique platform for engineering the BC in a non-volatile way and has high potential for optoelectronic and optospintronic applications.

## Methods

**Electronic structure calculation.** Our DFT calculations are performed using the projected augmented plane-wave method[40,41] as implemented in the Vienna *ab initio* simulation package (VASP).[42] The optimized atomic structure of the SnTe monolayer is obtained from the HSE functional.[37] The PBE functional of the generalized gradient approximation is used to describe the exchange-correlation interactions among electrons.[36] The isolated SnTe monolayer is considered within supercell geometries where the interlayer distance is 15 Å in the surface normal direction. The energy cutoff for the plane-wave-basis expansion is selected to be 450 eV. We used a 10×10×1 **k**-point grid to sample the entire Brillouin zone. The BC $\Omega(\mathbf{k})$ is calculated as follows:[1,43]

$$\Omega(\mathbf{k}) = -2\text{Im} \sum_n \sum_{n' \neq n} f_n \frac{\langle \psi_n(\mathbf{k})|v_x|\psi_{n'}(\mathbf{k})\rangle \langle \psi_{n'}(\mathbf{k})|v_y|\psi_n(\mathbf{k})\rangle}{(E_{n'}(\mathbf{k}) - E_n(\mathbf{k}))^2}, \tag{7}$$

where $n$ is the band index, $f_n$ is the Fermi-Dirac distribution function, $v_{x(y)}$ is the velocity operator, and $\psi_n(\mathbf{k})$ and $E_n(\mathbf{k})$ are the Bloch wave-function and energy, respectively, of the $n$-th band at point **k**. The BC and the spin BC are evaluated via the maximally localized Wannier function using the WANNIER90 package.[44,45] The intra-band BC and inter-band BC dipoles are estimated using a 2000×2000×1 **k**-point grid.

**Intra/inter-band BC dipoles.** The intra-band BC dipole, namely, $\mathbf{D}^{\text{intra}}$, is expressed as a function of the chemical potential $\mu$ as follows:

$$\mathbf{D}^{\text{intra}}(\mu) = \frac{1}{(2\pi)^2} \sum_n \int f(E_n(\mathbf{k}) - \mu) \nabla_\mathbf{k} \Omega_n(\mathbf{k}) \, d^2k, \tag{8}$$

where $\Omega_n$ represents the $n$-th band BC and $f(E_n(\mathbf{k}) - \mu)$ is the Fermi-Dirac distribution. The inter-band BC dipole $\mathbf{D}^{\text{inter}}$ is expressed as

$$\mathbf{D}^{\text{inter}}(\omega) = \frac{1}{(2\pi)^2} \int \Theta(\hbar\omega - \Delta E(\mathbf{k})) \nabla_{\mathbf{k}} \Omega_{vc}(\mathbf{k}) d^2 k. \tag{9}$$

Here, $\Omega_{vc} = i \sum_{v,c} \langle v|\partial_{k_x} H|c\rangle \langle c|\partial_{k_y} H|v\rangle / [\Delta E(\mathbf{k})]^2$ is the inter-band BC[20] between the valence and conduction bands [the summation running over each spin band of the valance ($v$) and the conduction ($c$) band], $\Delta E(\mathbf{k})$ is the inter-band transition energy between them, $\hbar\omega$ denotes the photon energy, and $\Theta$ is the Heaviside step function.

**Charge and spin photogalvanic current calculation.** To estimate the charge and spin currents induced by the circularly polarized light, the time-dependent DFT calculations are performed using a custom code based on the quantum ESPRESSO package.[46,47] The exchange and correlation interactions between electrons are described by the PBE-type generalized gradient approximation functional.[36] Norm-conserving pseudopotentials are used to describe the nuclei-electron interaction. In addition to the 10×10×1 grid points that are sampled by the Monkhorst-Pack scheme in the Brillouin zone, 16 time-reversal-symmetric **k**-points near the X valleys are employed to simulate the photo-excited currents more accurately. To investigate the real-time dynamics of the optical responses, circularly polarized light is applied in the velocity gauge $\mathbf{A}_{\pm}(t) = \frac{E_0}{\omega_0}(\mathbf{x}\sin(\omega_0 t) \pm \mathbf{y}\cos(\omega_0 t))$, where the electric field amplitude is $E_0 = 5.64 \times 10^{-3}$ V Å$^{-1}$ and the light frequency is set to the resonant direct band gap energy as $\hbar\omega_0 = 0.58$ eV. The charge and spin currents induced by the circularly polarized light are evaluated in terms of the expectation values of the spin and velocity operators as follows:[48]

$$\mathbf{J}(t) = \frac{-e}{V_2} \sum_{\mathbf{k}}^{BZ} \sum_n f_n \langle \psi_n(\mathbf{k},t)|\mathbf{v}|\psi_n(\mathbf{k},t)\rangle, \tag{10}$$

$$\mathbf{J}_{S_z}(t) = \frac{e}{2V_2} \sum_{\mathbf{k}}^{BZ} \sum_{n} f_n \langle \psi_n(\mathbf{k},t) | \{S_z, \boldsymbol{v}\} | \psi_n(\mathbf{k},t) \rangle. \qquad (11)$$

where $n$ is the band index, $f_n$ is the initial occupation of the Bloch state, and $V_2$ is the lattice surface area.

## Data Availability

The data that support the findings of this study are available from J.K. and K.W.K. upon reasonable request.

## Code Availability

The code that was used for the time-dependent DFT calculations is available from N.P. upon reasonable request.

## References

1   Thouless, D. J., Kohmoto, M., Nightingale, M. P. & den Nijs, M. Quantized Hall Conductance in a Two-Dimensional Periodic Potential. *Phys. Rev. Lett.* **49**, 405-408 (1982).

2   Xiao, D., Chang, M.-C. & Niu, Q. Berry phase effects on electronic properties. *Rev. Mod. Phys.* **82**, 1959-2007 (2010).

3   Yu, R. *et al.* Quantized Anomalous Hall Effect in Magnetic Topological Insulators. *Science* **329**, 61-64 (2010).

4   Fang, Z. *et al.* The Anomalous Hall Effect and Magnetic Monopoles in Momentum Space. *Science* **302**, 92-95 (2003).

5   Chang, M.-C. & Niu, Q. Berry Phase, Hyperorbits, and the Hofstadter Spectrum. *Phys. Rev. Lett.* **75**, 1348-1351 (1995).

6   Sundaram, G. & Niu, Q. Wave-packet dynamics in slowly perturbed crystals: Gradient corrections and Berry-phase effects. *Phys. Rev. B* **59**, 14915-14925 (1999).

7   Murakami, S., Nagaosa, N. & Zhang, S.-C. Dissipationless Quantum Spin Current at Room Temperature. *Science* **301**, 1348-1351 (2003).

8   Xiao, D., Yao, W. & Niu, Q. Valley-Contrasting Physics in Graphene: Magnetic Moment and Topological Transport. *Phys. Rev. Lett.* **99**, 236809 (2007).

9   Xiao, D., Liu, G.-B., Feng, W., Xu, X. & Yao, W. Coupled Spin and Valley Physics in Monolayers of MoS$_2$ and Other Group-VI Dichalcogenides. *Phys. Rev. Lett.* **108**, 196802 (2012).

10  Chang, C.-Z. *et al.* Experimental Observation of the Quantum Anomalous Hall Effect in a Magnetic Topological Insulator. *Science* **340**, 167-170 (2013).

11  Kim, J., Jhi, S.-H., MacDonald, A. H. & Wu, R. Ordering mechanism and quantum anomalous Hall effect of magnetically doped topological insulators. *Phys. Rev. B* **96**, 140410 (2017).

12  Nagaosa, N., Sinova, J., Onoda, S., MacDonald, A. H. & Ong, N. P. Anomalous Hall effect. *Rev. Mod. Phys.* **82**, 1539-1592 (2010).

13  Sinova, J., Valenzuela, S. O., Wunderlich, J., Back, C. H. & Jungwirth, T. Spin Hall effects. *Rev. Mod. Phys.* **87**, 1213-1260 (2015).

14  Berry, M. V. Quantal phase factors accompanying adiabatic changes. *Proc. R. Soc. London, Ser. A* **392**, 45-57 (1984).

15  Jungwirth, T., Niu, Q. & MacDonald, A. H. Anomalous Hall Effect in Ferromagnetic Semiconductors. *Phys. Rev. Lett.* **88**, 207208 (2002).

16  Deyo, E., Golub, L., Ivchenko, E. & Spivak, B. Semiclassical theory of the photogalvanic effect in non-centrosymmetric systems. Preprint at http://arxiv.org/abs/0904.1917 (2009).

17  Moore, J. E. & Orenstein, J. Confinement-Induced Berry Phase and Helicity-Dependent Photocurrents. *Phys. Rev. Lett.* **105**, 026805 (2010).

18   Sodemann, I. & Fu, L. Quantum Nonlinear Hall Effect Induced by Berry Curvature Dipole in Time-Reversal Invariant Materials. *Phys. Rev. Lett.* **115**, 216806 (2015).

19   Low, T., Jiang, Y., & Guinea, F. Topological currents in black phosphorus with broken inversion symmetry. *Phys. Rev. B.* **92**, 235447 (2015).

20   Xu, S.-Y. et al., Electrically switchable Berry curvature dipole in the monolayer topological insulator WTe$_2$. *Nat. Phys.* **14**, 900-906 (2018).

21   Ma, Q. et al. Observation of the nonlinear Hall effect under time-reversal-symmetric conditions. *Nature* **565**, 337–342 (2019).

22   Manna, K. *et al.* From Colossal to Zero: Controlling the Anomalous Hall Effect in Magnetic Heusler Compounds via Berry Curvature Design. *Phys. Rev. X* **8**, 041045 (2018).

23   de Juan, F., Grushin, A. G., Morimoto, T. & Moore, J. E. Quantized circular photogalvanic effect in Weyl semimetals. *Nat. Commun.* **8**, 15995 (2017).

24   Facio, J. I. *et al.* Strongly Enhanced Berry Dipole at Topological Phase Transitions in BiTeI. *Phys. Rev. Lett.* **121**, 246403 (2018).

25   Zhang, Y., Sun, Y. & Yan, B. Berry curvature dipole in Weyl semimetal materials: An ab initio study. *Phys. Rev. B* **97**, 041101 (2018).

26   Hosur, P. Circular photogalvanic effect on topological insulator surfaces: Berry-curvature-dependent response. *Phys. Rev. B* **83**, 035309 (2011).

27   Chang, K. *et al.* Discovery of robust in-plane ferroelectricity in atomic-thick SnTe. *Science* **353**, 274-278 (2016).

28   Mehboudi, M. *et al.* Two-dimensional disorder in black phosphorus and monochalcogenide monolayers. *Nano Lett.* **16**, 1704-1712 (2016).

29   Wu, M. & Zeng, X. C. Intrinsic ferroelasticity and/or multiferroicity in two-dimensional phosphorene and phosphorene analogues. *Nano Lett.* **16**, 3236-3241 (2016).

30   Fei, R., Kang, W. & Yang, L. Ferroelectricity and Phase Transitions in Monolayer Group-IV Monochalcogenides. *Phys. Rev. Lett.* **117**, 097601 (2016).

31   Xu, L., Yang, M., Wang, S. J. & Feng, Y. P. Electronic and optical properties of the monolayer group-IV monochalcogenides MX (M=Ge, Sn; X=S, Se, Te). *Phys. Rev. B* **95**, 235434 (2017).

32   Bychkov, Y. A. & Rashba, É. I. Properties of a 2D electron gas with lifted spectral degeneracy. *JETP Lett* **39**, 78 (1984).

33   Lee, H., Im, J. & Jin, H. Harnessing the giant in-plane Rashba effect and the nanoscale persistent spin helix via ferroelectricity in SnTe thin films. Preprint at http://arxiv.org/abs/1712.06112 (2017).

34   Kim, M., Im, J., Freeman, A. J., Ihm, J. & Jin, H. Switchable S=1/2 and J=1/2 Rashba bands in ferroelectric halide perovskites. *Proc. Natl. Acad. Sci.* **111**, 6900-6904 (2014).

35   Park, S. R., Kim, C. H., Yu, J., Han, J. H. & Kim, C. Orbital-Angular-Momentum Based Origin of Rashba-Type Surface Band Splitting. *Phys. Rev. Lett.* **107**, 156803 (2011).

36   Perdew, J. P., Burke, K. & Ernzerhof, M. Generalized Gradient Approximation Made Simple. *Phys. Rev. Lett.* **77**, 3865-3868 (1996).

37   Heyd, J., Scuseria, G. E. & Ernzerhof, M. Hybrid functionals based on a screened Coulomb potential. *J. Chem. Phys.* **118**, 8207-8215 (2003).

38   Cao, T. *et al.* Valley-selective circular dichroism of monolayer molybdenum disulphide. *Nat. Commun.* **3**, 887 (2012).

39   Sipe, J. E. & Shkrebtii, A. I. Second-order optical response in semiconductors. *Phys. Rev. B* **61**, 5337-5352 (2000).

40   Blöchl, P. E. Projector augmented-wave method. *Phys. Rev. B* **50**, 17953-17979 (1994).

41   Kresse, G. & Joubert, D. From ultrasoft pseudopotentials to the projector augmented-wave method. *Phys. Rev. B* **59**, 1758-1775 (1999).

42   Kresse, G. & Hafner, J. Ab initio molecular dynamics for liquid metals. *Phys. Rev. B* **47**, 558 (1993).

43   Yao, Y. *et al.* First Principles Calculation of Anomalous Hall Conductivity in Ferromagnetic bcc Fe. *Phys. Rev. Lett.* **92**, 037204 (2004).

44   Mostofi, A. A. *et al.* wannier90: A tool for obtaining maximally-localised Wannier functions. *Comput. Phys. Commun.* **178**, 685-699 (2008).

45   Kim, J. *et al.* Weyl node assisted conductivity switch in interfacial phase-change memory with van der Waals interfaces. *Phys. Rev. B* **96**, 235304 (2017).

46   Paolo, G. *et al.* QUANTUM ESPRESSO: a modular and open-source software project for quantum simulations of materials. *J. Phys: Condens. Matter* **21**, 395502 (2009).

47   Shin, D., Lee, G., Miyamoto, Y. & Park, N. Real-Time Propagation via Time-Dependent Density Functional Theory Plus the Hubbard U Potential for Electron–Atom Coupled Dynamics Involving Charge Transfer. *J. Chem. Theory Comput.* **12**, 201-208 (2016).

48   Shin, D. *et al.* Unraveling materials Berry curvature and Chern numbers from real-time evolution of Bloch states. *Proc. Natl. Acad. Sci.* **116**, 4135-4140 (2019).

**Acknowledgements** Financial support from the Basic Science Research Program of the National Research Foundation of Korea (NRF) under Grant No. 2016R1D1A1B03933255, 2017M3D1A1040828, 2019R1A2C1010498 (H.J.), and 2016R1D1A1B03931542, 2017R1A4A1015323 (D.S. and N.P.) is gratefully acknowledged. J. K. was supported by the NRF grant funded by the Korea government (MSIT) (No. 2019R1F1A1059743) and by Incheon National University Research Grant in 2019 (20190291). K.W.K. acknowledges financial support from the KIST Institutional Program and the National Research Council of Science & Technology (NST) (Grant No. CAP-16-01-KIST). K.W.K. and J.S. were also supported by the German Research Foundation (DFG) (No. SI 1720/2-1), the Alexander von Humboldt Foundation, and the Transregional Collaborative Research Center (SFB/TRR) 173 SPIN+X.

**Author Contributions** J.K. and S.H.L. performed first-principles calculations. D.S. and N.P. conducted the time-dependent DFT calculations. K.W.K. and J.S. constructed the model Hamiltonian and calculated the corresponding results. J.K., K.W.K. and H.J. analysed the data and wrote the paper with the help of the other authors.

**Competing interests** The authors declare no competing interests.

**Corresponding authors** Correspondence and requests for materials should be addressed to H.J. (email: hsjin@unist.ac.kr)

**Supplementary Information** is available in the online version of the paper.

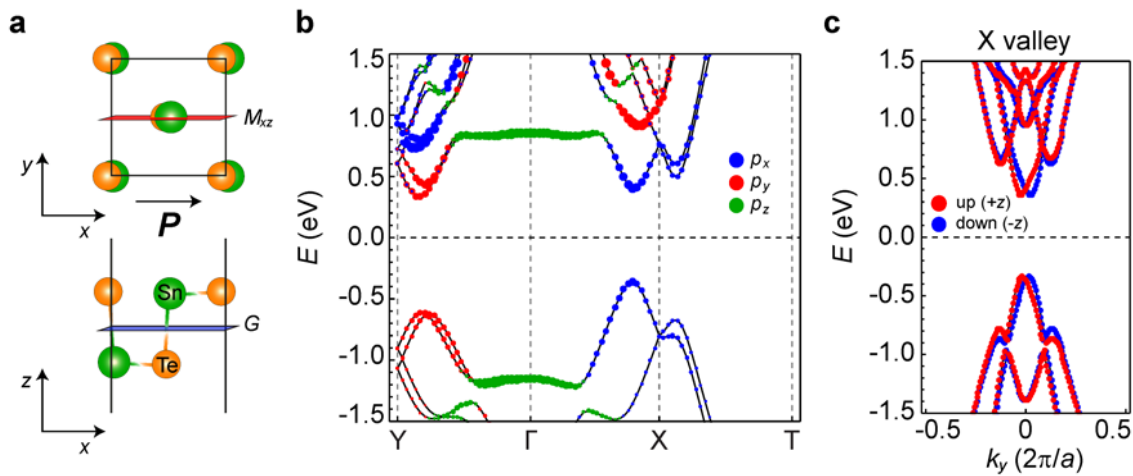

**Figure 1 | Atomic and electronic structure of the SnTe monolayer. a,** Top and side views of the SnTe monolayer. The ferroelectricity is induced by the in-plane atomic movements of the Sn (green spheres) and Te (orange spheres) atoms. $M_{xz}$ and $G$ denote the planes for the mirror symmetry and the gliding symmetry, respectively. **b,** The calculated band structure of the SnTe monolayer. The states near the Fermi levels are mainly composed of $p$ orbitals and their weights are plotted in three colours. **c,** The calculated band structure for the X valley. The band calculation is performed along the $k_y$-direction for the X valley. Spin-up (spin-down) states are marked by red (blue) dots. Due to the inversion symmetry breaking induced by the ferroelectricity, a Rashba-type unidirectional spin texture appears in the band structure.

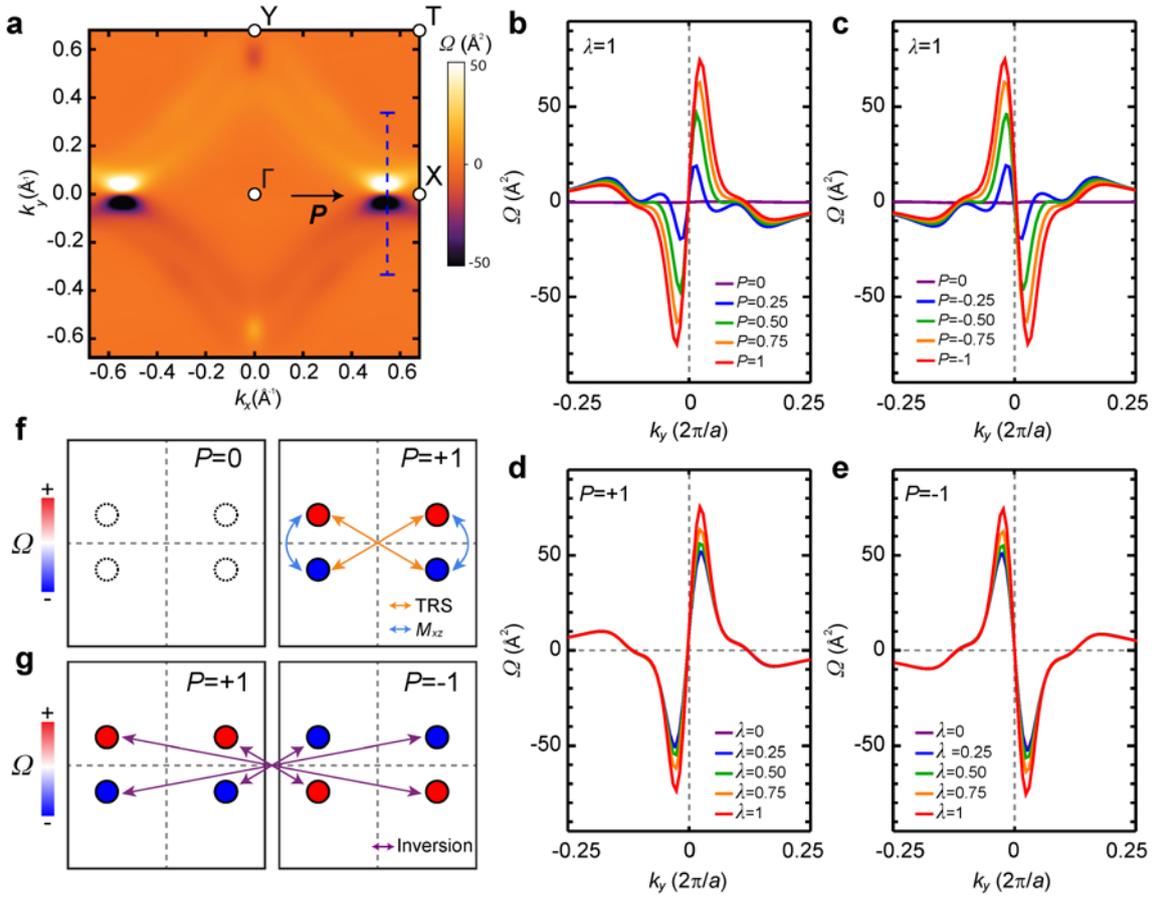

**Figure 2 | Berry curvature ($\Omega$) of the SnTe monolayer. a,** The calculated Berry curvature (BC) map in the first Brillouin zone. The BC dipoles are formed at the X valleys. **b-e,** The BC that is calculated along the vertical blue line in (**a**) under varying (**b, c**) ferroelectric polarization magnitude ($P$) and (**d, e**) spin-orbit coupling strength ($\lambda$), where the values of $P$ and $\lambda$ are normalized by the native values. The ferroelectricity reversal from (**b, d**) the positive to (**c, e**) the negative direction changes the sign of the BC distribution. These data imply that the BC dipole is strongly coupled with the ferroelectricity, rather than spin-orbit coupling. **f,** Schematic drawings of the BC according to the symmetry of the system. The red and blue circles denote the two opposite signs. The zero BCs are enforced by the combination of the time-reversal symmetry and the inversion symmetry in the absence of the ferroelectricity. When the in-plane ferroelectricity develops, the positive and negative BC peaks are formed along the $k_y$. The peaks that are connected by orange (blue) arrows denote time-reversal (mirror-symmetric) pairs; they

are transformed to each other by time-reversal (mirror) operation. **g,** Schematic diagrams of the response of the BC to the ferroelectric switching. Since the $+P$ and $-P$ configurations are inversion partners, the BCs of two configurations are opposite at the same $k$ point. The points that are connected by purple arrows denote inversion pairs.

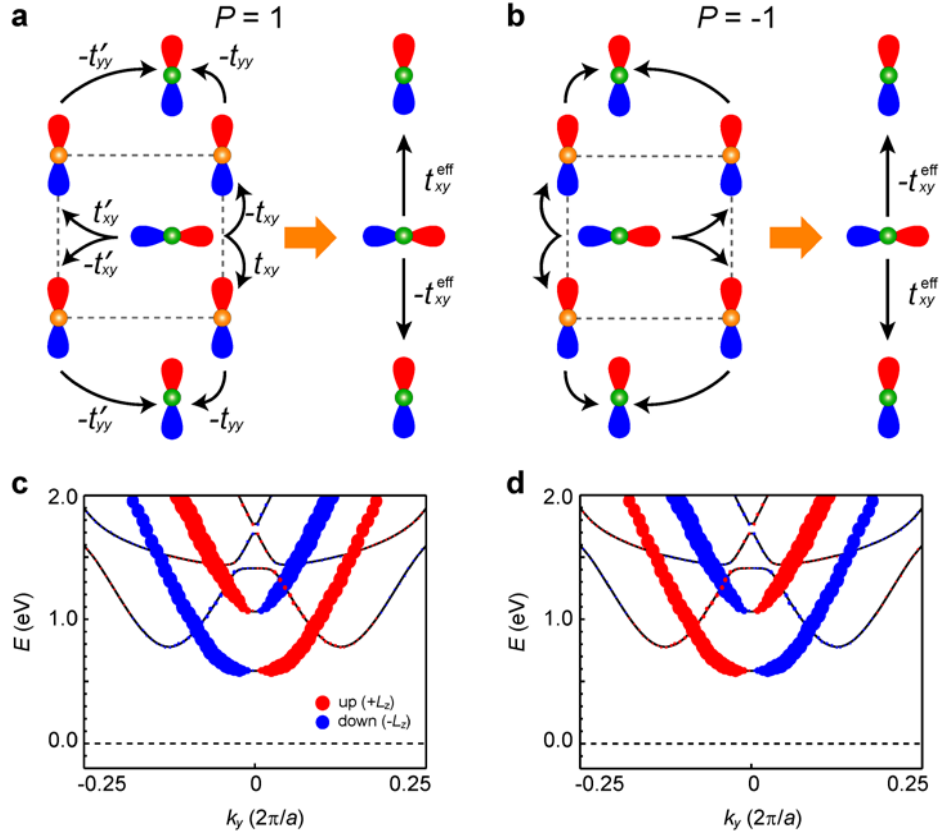

**Figure 3 | Ferroelectrically driven orbital Rashba effect. a-b**, The emergence of an asymmetric inter-orbital hopping that is induced by a ferroelectric polarization, which results in the orbital Rashba effect. An effective inter-orbital hopping ($t_{xy}^{\text{eff}} = \frac{t_{xy}t_{yy} - t'_{xy}t'_{yy}}{E_{\text{Sn}} - E_{\text{Te}}}$) from the Sn-$p_x$ orbital to the nearest-neighbour Sn-$p_y$ orbital along the *y*-axis is allowed by the ferroelectric displacement. The sign of the effective hopping is determined by the polarization direction. The dumbbells represent $p_x$ and $p_y$ orbitals of Sn and Te atoms, whose with the atomic energies are $E_{\text{Sn}}$ and $E_{\text{Te}}$, respectively; the blue (red) coloured region means a positive (negative) value of the *p* orbital wave function. Green balls, Sn; Orange balls, Te. **c-d**, The orbital angular momentum texture that was obtained via density functional theory calculations in the absence of spin-orbit coupling depends on the polarization direction, which supports the presence of the ferroelectrically coupled orbital Rashba effect. As predicted in our analytic model, the orbital angular momentum is an odd function with respect to $k_y$ and the orbital-split-

off states show opposite signs at the same $k$ point. From the close correlation between Berry curvature (BC) and orbital angular momentum, the orbital Rashba effect can bring a similar BC distribution, possibly leading to the BC dipole along the $k_y$-direction.

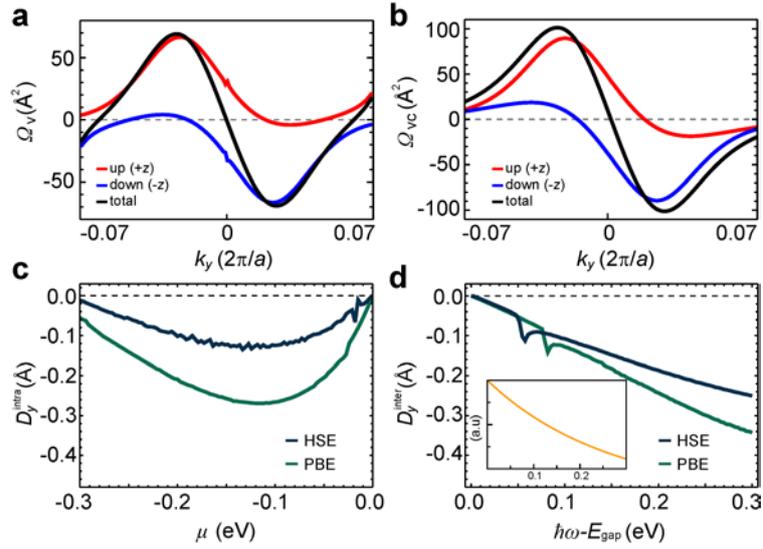

**Figure 4 | Berry curvature and Berry curvature dipoles. a,** The calculated Berry curvatures (BCs) for the highest valence bands near the X valley. The spin-up (spin-down) component is represented by a red (blue) line. **b,** The calculated inter-band BCs between the highest valence bands and the lowest conduction bands near the X valley. The spin-up (spin-down) component is represented by a red (blue) line. **c,** The intra-band BC dipole as a function of the Fermi level and **d,** the inter-band BC dipole as a function of the photon frequency, which were calculated from the PBE and HSE functionals. The inset shows the inter-band BC dipole calculated from our analytic formalism, Eq. (6), in which the unit of the horizontal axis is the same as the numerical result and the vertical axis of the inter-band BC dipole is presented in arbitrary unit.

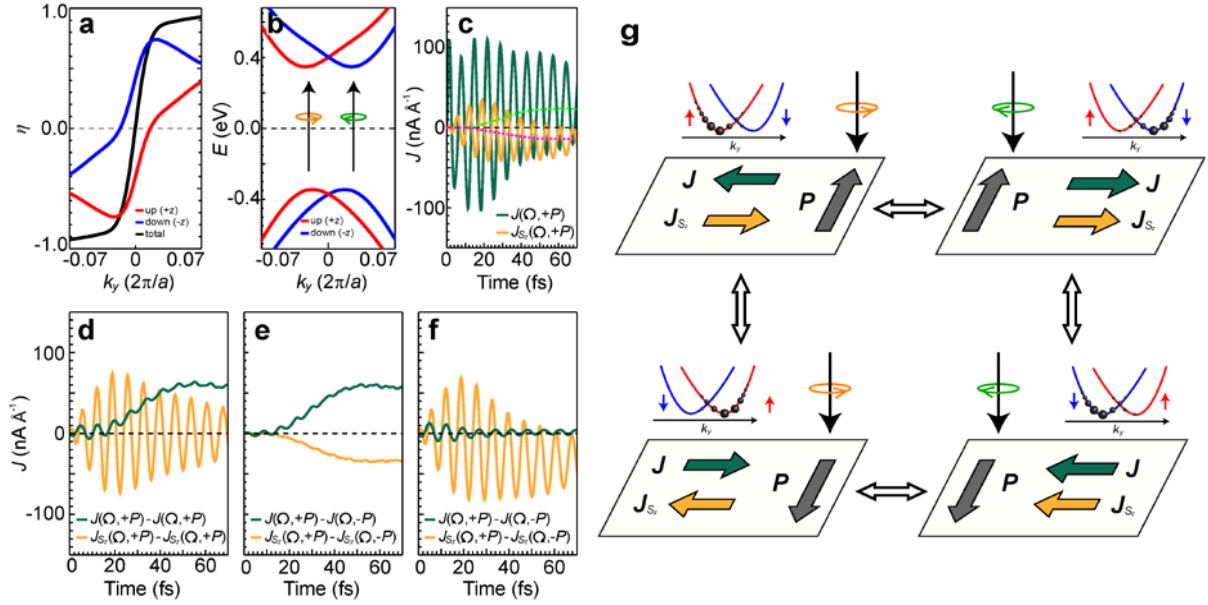

**Figure 5 | Switchable charge and spin photogalvanic effects. a,** The calculated circular dichroism ($\eta$) for the band-edge transitions near the X valley. The adsorption of the left-handed (right-handed) light prevails when $\eta$ is greater (smaller) than zero. **b,** Spin-selective circular dichroic excitation near the X valley. The spin-up (spin-down) states are marked by red (blue) lines. An orange (green) arrow represents the right-handed (left-handed) light. **c,** Charge (green line) and spin (yellow lies) currents under the right-handed light, which were calculated by the time-dependent density functional theory method. The dashed guidelines (cyan and magenta) were obtained by averaging out the rapidly oscillating contributions and represent the charge and spin photocurrents, respectively. **d-f,** The calculated charge (green line) and spin (yellow lines) current differences that are caused (d) by the photon helicity, (e) by the ferroelectric polarization, and (f) by both the photon helicity and the ferroelectric polarization. Here, the spin current in (d) and charge current in (e) gives no contribution after averaging out the oscillating contributions. This observation implies that the reversal of the photon helicity changes the direction of the charge current only, while the reversal of the ferroelectric polarization reverses the direction of both charge and spin photocurrents. **g,** Schematic drawings of the circular photogalvanic effect being controlled by the helicity of

the circularly polarized light and the ferroelectric polarization direction; the observations in (c)-(f) for four different combinations of the photon helicity and the ferroelectric polarization are depicted. In this way, one can independently control the direction of the charge and spin currents induced by the incident circularly polarized light. Here, the inversion and time-reversal operations are denoted by the vertical and the horizontal white arrows, respectively.

# Supplementary Information for "Prediction of ferroelectricity-driven Berry curvature enabling charge- and spin-controllable photocurrent in tin telluride monolayers"

J. Kim *et al.*

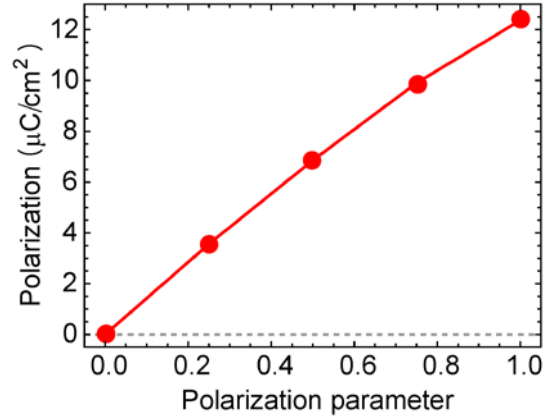

**Supplementary Figure 1 | Evolution of spontaneous electric polarization.** The calculated electric polarization of the SnTe monolayer is increased with the polarization parameter normalized by the native value along the *x*-axis.

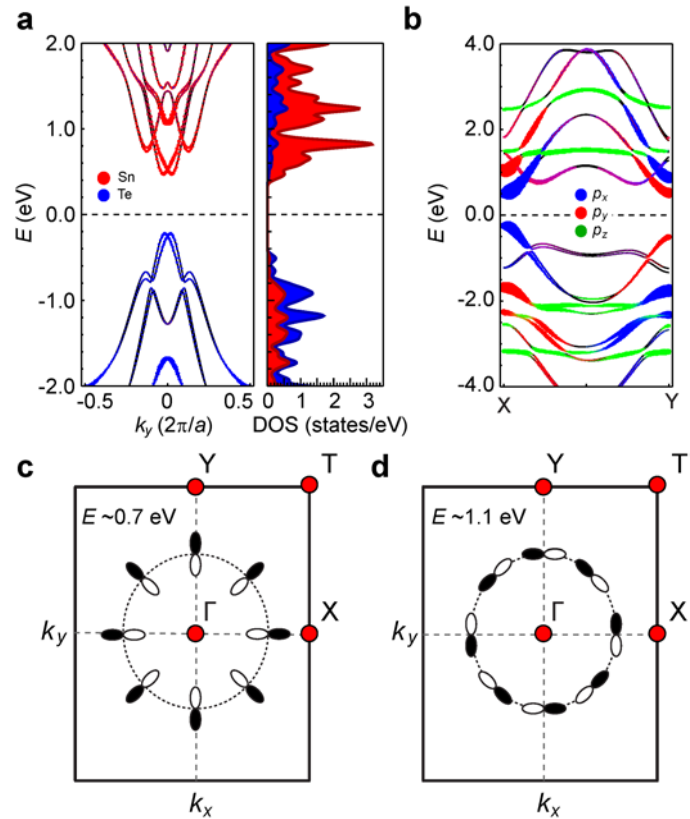

**Supplementary Figure 2 | Electronic structure of SnTe. a**, The atom-projected band structure and density of states (DOS) near the X valley. The contribution of Sn (Te) is represented in red (blue). **b**, The orbital-projected band structure along the circular path from the X valley to the Y valley. The contributions of $p_x$, $p_y$, and $p_z$ orbitals are represented in blue, red, and green, respectively. **c-d,** The atomic orbital projection of (**c**) the lowest (~0.7

eV) and (**d**) the next lowest (~1.1 eV) conduction bands along the circular line. Special points (Γ, X, Y, T) are denoted by red dots.

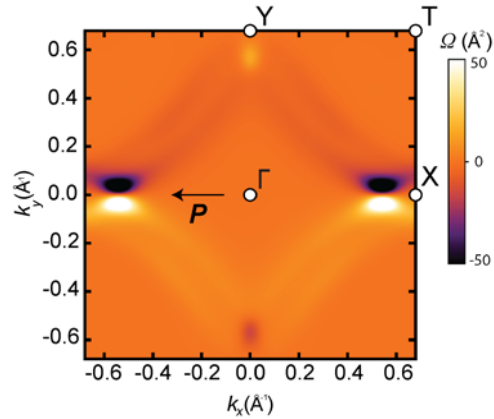

**Supplementary Figure 3 | Berry curvature ($\Omega$) of the SnTe monolayer with the opposite polarization direction.** The **c**alculated Berry curvature (BC) map in the first Brillouin zone. The BC dipoles are formed at the X valleys.

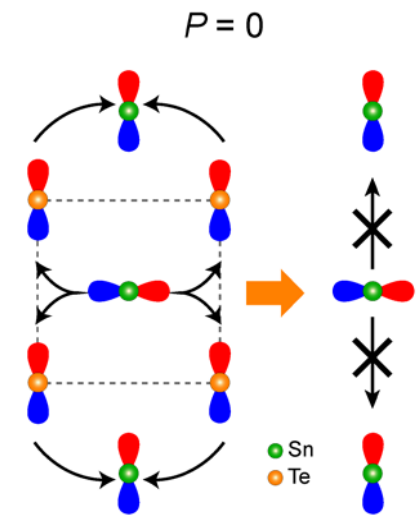

**Supplementary Figure 4 | Schematic drawing of the inter-orbital hopping channel in the SnTe monolayer.** The inter-orbital hopping between Sn atoms is absence without ferroelectric polarization. The dumbbells represent $p_x$ and $p_y$ orbitals of Sn and Te atoms; the blue (red) coloured region means a positive (negative) value of the $p$ orbital wave function.

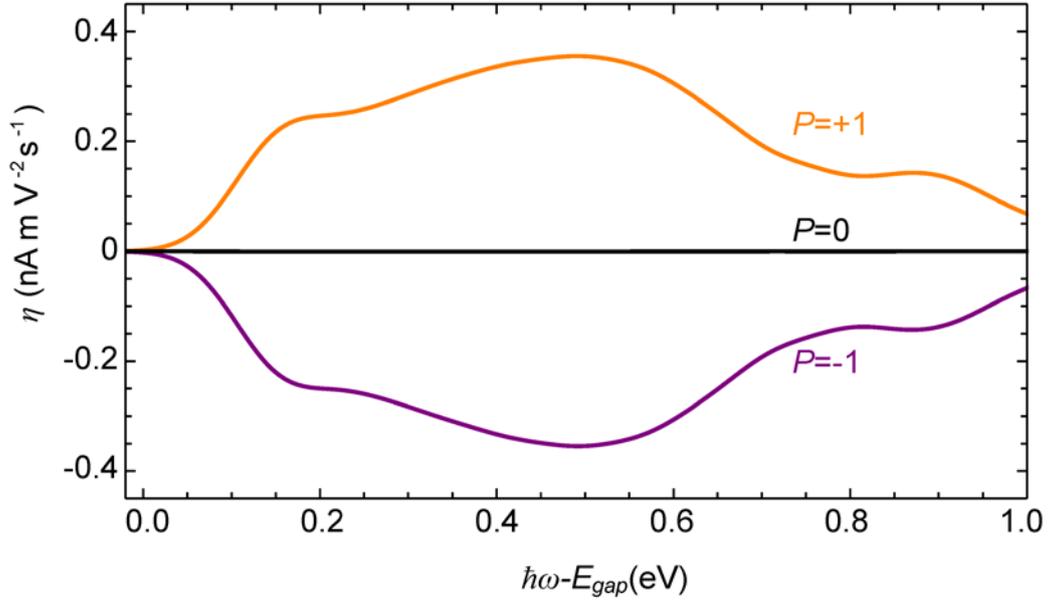

**Supplementary Figure 5 | Calculated spectral dependence of the photocurrent in the SnTe monolayer.** We calculate $\eta$ by using the formalism in Ref. 2. The photocurrent is then determined by $dJ_y/dt = \eta E_x(\omega)E_y(-\omega)$. Note that the photocurrent is reversed by the polarization switching from $P=+1$ to $P=-1$. Due to the reduced dimensionality of the SnTe monolayer, the units of $J$ and $\eta$ are changed accordingly in consideration of the 10 Å thickness of the vacuum layer.

**Supplementary Note 1 | Tight-binding derivation of the orbital Rashba effect.** From the atomic structure of the SnTe monolayer as shown in Fig. 1a, we utilize the gliding symmetry of the system in constructing the tight-binding model. The system is invariant under the gliding operators $T(a\mathbf{x}')v_x$ and $T(a\mathbf{y}')v_x$, where $T$ is the translational operator, $\mathbf{x}' = (\mathbf{x}+\mathbf{y})/\sqrt{2}$, $\mathbf{y}' = (-\mathbf{x}+\mathbf{y})/\sqrt{2}$, and $v_i$ (for $i = x, y, z$) is the Pauli matrix in Sn/Te space. The simultaneous eigenstates of these operators are

$$|\mathbf{k}, A, \pm\rangle = \frac{1}{\sqrt{N}} \sum_{nm} e^{i(nk_{x'}+mk_{y'})a} (\pm)^{n+m} |n, m, A\rangle, \quad (1)$$

where $|n, m, A\rangle$ is the electronic state localized at A=Sn, Te atom at the position $na\mathbf{x}' + ma\mathbf{y}'$, $|\mathbf{k}, A, \pm\rangle$ is the corresponding Bloch state, and $N$ is the total number of sites. Here $\pm$ degree of freedom comes from different eigenvalues for the gliding operators. Each Bloch state has 4 degrees of freedom: 2 for Sn/Te atomic sites and 2 for $p_x$/$p_y$-orbitals. For simplicity, we discard the spin degree of freedom, but the derivation keeps the same for spin cases. Below we write the Bloch state in terms of the momentum in the $x, y$ coordinate, by using $k_{x,y} = (k_{x'} \mp k_{y'})/\sqrt{2}$ and $a' = \sqrt{2}a$.

We note that $|\mathbf{k}, A, +\rangle = |\mathbf{k} + (2\pi/a')\mathbf{y}, A, -\rangle$, and thus the construction of the Hamiltonian only for + is sufficient for our purpose since the result for – is straightforwardly given by replacing $\mathbf{k}$ by $\mathbf{k} + (2\pi/a')\mathbf{y}$. The final result does not change with this choice.

As considering the nearest-neighbour hopping integrals, the hopping within a unit cell is given by $H_{\text{hop},1} = -t_\pi \sum_{nm} \mathbf{c}_{n,m}^\dagger v_x \mathbf{c}_{n,m}$ because the hopping to the $z$ direction is always composed of π-bonding. Here $\mathbf{c}_{n,m}^\dagger = \left(c_{n,m,\text{Sn},p_x}^\dagger, c_{n,m,\text{Sn},p_y}^\dagger, c_{n,m,\text{Te},p_x}^\dagger, c_{n,m,\text{Te},p_y}^\dagger\right)$, $\mathbf{c}_{n,m}$ is its conjugate transpose, and $c_{n,m,A,p_i}^\dagger$ is the creation operator for electrons localized at A atom at the site $(n, m)$. We define the creation operators for the Bloch states $c_{\mathbf{k},A,p_i}^\dagger$ similarly.

The hopping processes to neighbouring unit cells are more complicated. If the two atoms are distant along $\mathbf{r}_1$ direction, the hopping integral of $p_{r_1}$ orbitals is $t_\sigma$ while the hopping integral of $p_{r_2}$ is $t_\pi$ where $\mathbf{r}_2 = \mathbf{z} \times \mathbf{r}_1$. Then the hopping term is given in the form of $t_\sigma |p_{r_1}\rangle\langle p_{r_1}| + t_\pi |p_{r_2}\rangle\langle p_{r_2}|$. Defining $\theta_\mathbf{v} = \arg(v_x + iv_y)$ for an arbitrary vector $\mathbf{v}$, $|p_{r_1}\rangle = \cos\theta_{\mathbf{r}_1}|p_x\rangle + \sin\theta_{\mathbf{r}_1}|p_y\rangle$ and $|p_{r_2}\rangle = \cos\theta_{\mathbf{r}_2}|p_x\rangle + \sin\theta_{\mathbf{r}_2}|p_y\rangle$. Then, the hopping term is written in $(|p_x\rangle, |p_y\rangle)$ basis and corresponds to the following matrix.

$$R(\theta_{\hat{\mathbf{r}}_1}) \begin{pmatrix} t_\sigma & 0 \\ 0 & t_\pi \end{pmatrix} R(-\theta_{\hat{\mathbf{r}}_1}), \tag{2}$$

where $R(\theta)$ is the rotational matrix. For instance, the hopping term from $|n, m, \text{Sn}\rangle$ to $|n+1, m, \text{Te}\rangle$ in Fig. 3a is

$$\begin{pmatrix} c^\dagger_{n,m,\text{Te},p_x} & c^\dagger_{n,m,\text{Te},p_y} \end{pmatrix} R(\theta_{a\mathbf{x}'-d\mathbf{x}}) \begin{pmatrix} t_\sigma & 0 \\ 0 & t_\pi \end{pmatrix} R(-\theta_{a\hat{\mathbf{x}}'-d\hat{\mathbf{x}}}) \begin{pmatrix} c_{n,m,\text{Sn},p_x} \\ c_{n,m,\text{Sn},p_y} \end{pmatrix}, \tag{3}$$

where $d$ is the atomic displacement of Sn atoms, which is the ferroelectric polarization parameter. One realizes that, for $d = 0$, $\theta_{a\mathbf{x}'-d\mathbf{x}} = \pi/4$ thus the hopping integrals for both $p_x$ and $p_y$ orbitals are both $(t_\sigma + t_\pi)/2$. However, for $d \neq 0$, the hopping integrals are different (Fig. 3a and Supplementary Fig. 4). We denote the collection of the all possible hopping terms by $H_{\text{hop},2}$ without presenting its complicated form here. The periodicity of the system guarantees that $H_{\text{hop}} = H_{\text{hop},1} + H_{\text{hop},2}$ is diagonal in $\mathbf{k}$. Note also that the magnitude of the hopping integrals are different for hoppings to the $x$ direction and those to the $-x$ direction due to different distances (see the difference between $t'_{xy}$ and $t_{xy}$ in Fig. 3a). We denote the difference hopping integrals by $t_{\sigma/\pi}$ and $t'_{\sigma/\pi}$ accordingly. Their difference is proportional to the ferroelectric polarization parameter $d$, and thus we define $t_{\sigma/\pi} - t'_{\sigma/\pi} \equiv dt_{\sigma/\pi,\Delta}$.

We take the simplest on-site term given by

$$H_{\text{on-site}}(\mathbf{k}) = \sum_{\substack{i=x,y \\ A=\text{Sn,Te}}} E_A(\mathbf{k}) c^\dagger_{\mathbf{k},A,p_i} c_{\mathbf{k},A,p_i}. \tag{4}$$

It is also possible to generalize this model. For instance, one may include spin-orbit coupling

(SOC) to obtain **k**-dependent SOC parameter introduced in the main text. After tedious algebra, we obtain the following total Hamiltonian $H_{\text{on-site}} + H_{\text{hop}}$.

$$H(\mathbf{k}) = \sum_{\substack{i=x,y \\ A=\text{Sn,Te}}} c^\dagger_{\mathbf{k},A,p_i} h(\mathbf{k}) c_{\mathbf{k},A,p_i} \tag{5}$$

$$\begin{aligned}
h(\mathbf{k}) = & \begin{pmatrix} E_{\text{Sn}}(\mathbf{k}) & 0 \\ 0 & E_{\text{Te}}(\mathbf{k}) \end{pmatrix} \\
& - \left[ t_\pi + 2(t_\sigma + t_\pi) \cos\frac{k_x a}{\sqrt{2}} \cos\frac{k_y a}{\sqrt{2}} - 2(t_\sigma - t_\pi)\tau_x \sin\frac{k_x a}{\sqrt{2}} \sin\frac{k_y a}{\sqrt{2}} \right] \nu_x \\
& - \left[ 2d(t_{\sigma,\Delta} + t_{\pi,\Delta}) \sin\frac{k_x a}{\sqrt{2}} \cos\frac{k_y a}{\sqrt{2}} + 2d(t_{\sigma,\Delta} - t_{\pi,\Delta})\tau_x \cos\frac{k_x a}{\sqrt{2}} \sin\frac{k_y a}{\sqrt{2}} \right. \\
& \left. - \frac{2\sqrt{2}d}{a}(t_\sigma - t_\pi) \sin\frac{k_x a}{\sqrt{2}} \cos\frac{k_y a}{\sqrt{2}} \tau_z \right] \nu_y.
\end{aligned} \tag{6}$$

The total Hamiltonian is not diagonal in Sn/Te space. To obtain an effective Hamiltonian for Sn only, we take the Schrieffer-Wolff transformation: $H' = e^S(H_{\text{on-site}} + H_{\text{hop}})e^{-S}$, where $S$ is an anti-hermitian operator. In our case, the choice $S = \nu_z H_{\text{hop}}/[E_{\text{Sn}}(\mathbf{k}) - E_{\text{Te}}(\mathbf{k})]$ satisfies $[H_{\text{on-site}}, S] = H_{\text{hop}}$. Then, the Baker-Hausdorff lemma gives $H' = H_{\text{on-site}} + H_{\text{hop,eff}}$ where $H_{\text{hop,eff}} = [S, H_{\text{hop}}]/2$ is a second order correction of hoppings. Projecting $H_{\text{hop,eff}}$ to Sn gives an effective Hamiltonian for Sn. We denote the resulting $2 \times 2$ matrix by $H_{\text{hop,eff,Sn}}$.

After a long algebra, the $\tau_y$ component of $H_{\text{hop,eff,Sn}}$ up to first order in $d$ is

$$\frac{1}{2}\text{Tr}[\tau_y H_{\text{hop,eff,Sn}}] = \frac{8d}{a'} \frac{(t_\sigma - t_\pi)^2}{E_{\text{Sn}}(\mathbf{k}) - E_{\text{Te}}(\mathbf{k})} \sin^2\frac{k_x a'}{2} \sin k_y a'. \tag{7}$$

Near the X valley, $k_x \approx k_X$ is mostly constant, thus we obtain the orbital Rashba term proportional to the ferroelectric parameter $d$.

To consider the SOC energy in the on-site term, one may add

$$H_{\text{on-site,SOC}}(\mathbf{k}) = \sum_{A=\text{Sn,Te}} \lambda_A \begin{pmatrix} c^\dagger_{\mathbf{k},A,p_x} & c^\dagger_{\mathbf{k},A,p_y} \end{pmatrix} \tau_y \sigma_z \begin{pmatrix} c_{\mathbf{k},A,p_x} \\ c_{\mathbf{k},A,p_x} \end{pmatrix}. \tag{8}$$

By applying the same Schrieffer-Wolff transformation, we obtain the following SOC term:

$$-\left\{\frac{8(t_\sigma-t_\pi)^2\bar{\lambda}}{(E_{\text{Sn}}-E_{\text{Te}})^2}\sin^2\frac{k_xa'}{2}\sin^2\frac{k_ya'}{2}+\frac{2\Delta\lambda}{(E_{\text{Sn}}-E_{\text{Te}})^2}\left[t_\pi+2(t_\sigma+t_\pi)\cos\frac{k_xa'}{2}\cos\frac{k_ya'}{2}\right]^2\right\}\tau_y\sigma_z, \quad (9)$$

which contribute to $H_{\text{hop,eff,Sn}}$ as an effective SOC energy in the form of $H_{\text{SOC}}(\mathbf{k})=\lambda_{\mathbf{k}}\tau_y\sigma_z$ as shown in the main text. Here $\bar{\lambda}=(\lambda_{\text{Sn}}+\lambda_{\text{Te}})/2$ and $\Delta\lambda=(\lambda_{\text{Sn}}-\lambda_{\text{Te}})/2$.

**Supplementary Note 2 | Comparison between the analytic model and DFT calculations.**

We start from the model Hamiltonian for the conduction Sn-$p_x$ and Sn-$p_y$ bands near the X valley.

$$H(\mathbf{k})=H_0(\mathbf{k})+H_{\text{SOC}}(\mathbf{k})+H_{\text{FE}}(\mathbf{k}), \quad (10)$$

$$H_0(\mathbf{k})=E_X(\mathbf{k})-J\cos 2\theta_{\mathbf{k}}\tau_z-J\sin 2\theta_{\mathbf{k}}\tau_x, \quad (11)$$

$$H_{\text{SOC}}(\mathbf{k})=\frac{2\lambda_{\mathbf{k}}}{\hbar^2}\mathbf{L}\cdot\mathbf{S}, \quad (12)$$

$$H_{\text{FE}}(\mathbf{k})=\alpha_Lk_yL_z=|\alpha_L|\mathbf{L}\cdot(\widehat{\mathbf{P}}\times\mathbf{k}). \quad (13)$$

where, in addition to the other terms introduced in the main text, we introduce the SOC Hamiltonian $H_{\text{SOC}}(\mathbf{k})$ to reproduce also the Rashba type band structure presented in Fig. 1c. Here, $\lambda_{\mathbf{k}}$ is the effective SOC strength for Sn atoms and $\mathbf{S}$ is the spin angular momentum operator. The effective SOC contains the $\mathbf{k}$ dependence as we integrate out the Te $p$-orbital degrees of freedom as explicitly shown above. Within the $p_x$, $p_y$-orbital subspace, $H_{\text{SOC}}(\mathbf{k})=\lambda_{\mathbf{k}}\sigma_z\tau_y$, where $\sigma_i$ is the spin Pauli matrix.

Keeping the first-order terms in $\lambda_{\mathbf{k}}$, the energy eigenvalues of $H(\mathbf{k})$ are given by

$$E_{\mathbf{k}n\sigma}=E_X(\mathbf{k})+(-1)^n\left(J_{\mathbf{k}}+\sigma\frac{\lambda_{\mathbf{k}}\alpha_L\hbar}{J_{\mathbf{k}}}k_y\right), \quad (14)$$

where $n(=1,2)$ and $\sigma(=\pm)$ are the orbital and spin indices, respectively, and $J_{\mathbf{k}}=\sqrt{J^2+(\alpha_L\hbar k_y)^2}$ is the modified orbital splitting. The last term in Supplementary Eq. (14) represents a $k_y$-linear spin splitting, which is the unidirectional Rashba spin splitting that is shown in Fig. 1c.[33] The spin texture calculated from $\langle\psi_{n\sigma}(\mathbf{k})|\mathbf{S}|\psi_{n\sigma}(\mathbf{k})\rangle = \sigma(\hbar/2)\mathbf{z}$ also accords with our DFT calculations.

The BC of the lowest energy conduction band $(n = 1)$ is obtained as

$$\Omega(\mathbf{k}) = \frac{2\alpha_L J^2}{J_\mathbf{k}^3 \hbar}\partial_{k_x}\theta_\mathbf{k}, \tag{15}$$

which reproduces the main features of the BC dipole presented in Fig. 2a. From the linear dependence of the BC on the orbital Rashba coefficient $\alpha_L$, $\Omega(\mathbf{k})$ is (i) induced by the ferroelectric polarization and (ii) switchable by reversing $P$. The BC is also (iii) independent of $\lambda_\mathbf{k}$ (up to first order), (iv) an odd function of $k_y$, and (v) an even function of $k_x$.

The orbital Rashba effect of Supplementary Eq. (13) leads to a finite expectation value of the *z*-component orbital angular momentum, which is zero without the ferroelectrically induced anti-symmetric hopping (Fig. 3a, b). The orbital angular momentum texture from our analytic model is written as

$$\langle\psi_{n\sigma}(\mathbf{k})|\mathbf{L}|\psi_{n\sigma}(\mathbf{k})\rangle = (-1)^n\left[\frac{\alpha_L\hbar^2}{J_\mathbf{k}}k_y + \sigma\frac{\lambda_\mathbf{k}\hbar}{J_\mathbf{k}}\left(1 - \frac{\alpha_L^2\hbar^2}{J_\mathbf{k}^2}k_y^2\right)\right]\mathbf{z}. \tag{16}$$

The orbital angular momentum that is evaluated from the DFT calculation without SOC (Fig. 3c, d) is consistent with the first term in Supplementary Eq. (16); the orbital angular momentum texture (i) is odd in $k_y$ (thus, it changes its sign at $k_y = 0$ within a single band), (ii) has the opposite signs for *n* =1 and 2, and (iii) changes its sign by reversing *P*. Based on the agreement with the DFT results, our minimal model demonstrates that the intriguing BC structure of the SnTe monolayer is developed via the orbital Rashba effect that originates from the in-plane ferroelectricity.

As a side remark, we confirmed that extending our model by including all the *p* orbitals does not alter our results in Supplementary Eqs. (14)–(16), up to first order in $\lambda_\mathbf{k}$.

**Supplementary Note 3 | Spin Berry curvature distribution.** To support the validity of our minimal model presented in the main text, we additionally calculate the spin-resolved BC distribution and compare it with the DFT calculations.

Supplementary Fig. 6 shows the DFT calculation for the spin BC, given by

$$\Omega_s(\mathbf{k}) = -\frac{\hbar}{2}\text{Im}\sum_{n}\sum_{n'\neq n} f_n \frac{\langle\psi_n(\mathbf{k})|\{\sigma_z,v_x\}|\psi_{n'}(\mathbf{k})\rangle\langle\psi_{n'}(\mathbf{k})|v_y|\psi_n(\mathbf{k})\rangle}{(E_{n'}(\mathbf{k})-E_n(\mathbf{k}))^2}. \tag{17}$$

Similar to the BC distribution, the spin BC is mainly concentrated near the X valley. Different from the BC whose total sum over the Brillouin zone is vanishing due to the time-reversal symmetry, the spin BC gives a non-zero net flux. This implies that a large spin Hall conductivity can be acquired when the SnTe monolayer is slightly doped.

A distinction between $\Omega$ and $\Omega_s$ can also be found in their dependence on ferroelectricity and SOC. While the BC varies in proportion to $P$, the spin BC remains large irrespective of $P$ (Supplementary Fig. 6b). Moreover, the overall sign of the spin BC is independent of the polarization direction. Meanwhile, the spin BC drastically changes with varying $\lambda$ (Supplementary Fig. 6c); in contrast to the BC, the spin BC scales with $\lambda$ and disappears as $\lambda$ goes to zero (Supplementary Fig. 6d). Our results indicate that the SOC is the key ingredient for the spin BC while the ferroelectricity is less relevant.

The symmetry argument presented in the main text can also be applicable to the spin BC, $\Omega_s(\mathbf{k}) = \Omega_+(\mathbf{k}) - \Omega_-(\mathbf{k})$. The time reversal symmetry and the mirror symmetry of the system imply $\Omega_s(\mathbf{k}) = \Omega_s(-\mathbf{k})$ and $\Omega_s(k_x, k_y) = \Omega_s(k_x, -k_y)$ respectively. Without SOC, the two spin channels become identical and thus $\Omega_+(\mathbf{k}) = \Omega_-(\mathbf{k})$. Consequently, $\Omega_s(\mathbf{k})$ is zero without $\lambda$ (Supplementary Fig. 6d). Once we consider the ferroelectric reversal, $\Omega_\pm^{+P}(\mathbf{k}) = \Omega_\pm^{-P}(-\mathbf{k})$ implies $\Omega_s^{+P}(\mathbf{k}) = \Omega_s^{-P}(\mathbf{k})$ (Supplementary Fig. 6e). Therefore, the symmetry argument well explains the DFT results shown in Supplementary Fig. 6a-c.

Lastly, we compare the DFT calculation with our analytic model. From the minimal model presented in the main text, we obtain

$$\Omega_s(\mathbf{k}) = \frac{J^2}{J_\mathbf{k}^3 \hbar}(\nabla_\mathbf{k}\theta_\mathbf{k} \times \nabla_\mathbf{k}\lambda_\mathbf{k})_z - \frac{3\hbar\lambda_\mathbf{k}\alpha_L^2 J^2 k_y}{J_\mathbf{k}^5}\partial_{k_x}\theta_\mathbf{k}, \tag{18}$$

which reproduces the spin BC distribution presented in Supplementary Fig. 6 well: (i) dependent on $\lambda_\mathbf{k}$, (ii) independent of the sign of $P$, and (iii) an even function of $k_x$ and $k_y$. The first term in Supplementary Eq. (18) is independent of the ferroelectric polarization (up to first order), which explains the non-vanishing $\Omega_s(\mathbf{k})$ at $P = 0$ ($\alpha_L = 0$) in Supplementary Fig. 6b. The second term is the ferroelectric contribution in the spin BC, originating from both the orbital Rashba effect and the SOC.

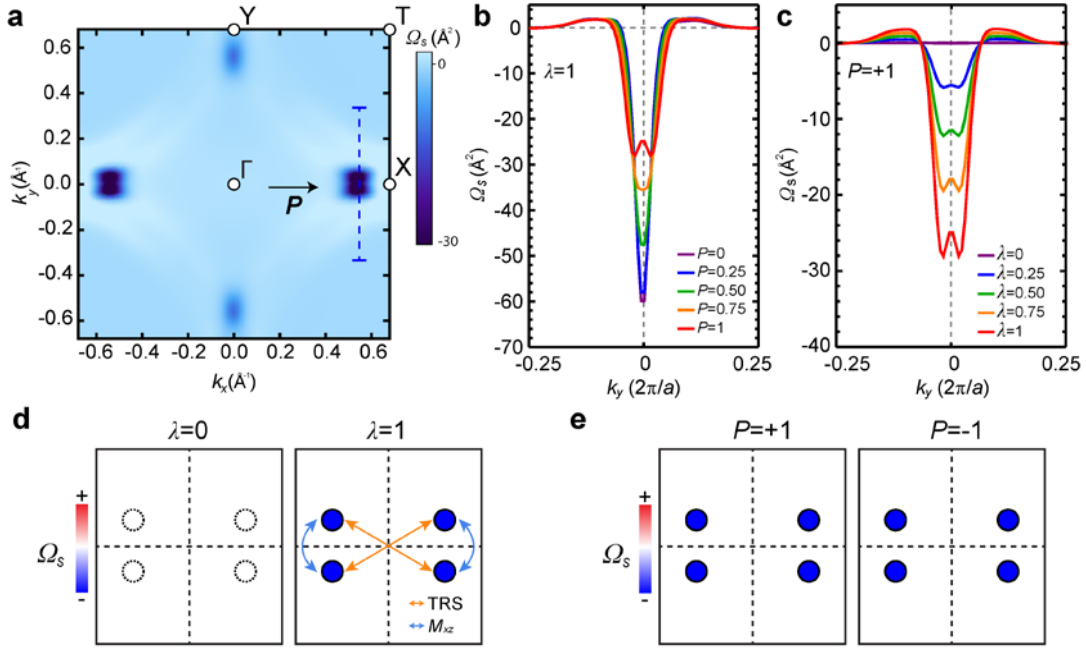

**Supplementary Figure 6 | spin Berry curvature ($\Omega_s$) of SnTe monolayer. a,** Calculated spin Berry curvature (BC) map in the first Brillouin zone. The spin BC monopole is formed at the X valleys. **b-c,** the spin BC calculated along the vertical blue line in (**a**) for varying the magnitude of (**b**) the ferroelectric polarization ($P$) and (**c**) the SOC strength ($\lambda$), where the values of $P$ and $\lambda$ are normalized by the native values. **d,** Schematic drawings for the spin BC following the symmetry of the system. **e,** Schematic drawings for the response of the spin BC to the ferroelectric switching.

**Supplementary Note 4 | Calculation of the optoelectronic responses using the analytic theory.** To analytically calculate the optoelectronic responses, we start with the Hamiltonian in Supplementary Eq. (10) and calculate the charge and spin photogalvanic effect.

According to Ref. 1, the charge current generated by the right/left-circularly polarized light is given by

$$J_{y,\pm} = \pm \frac{2\pi e^3 \tau E_0^2}{\hbar^2} D^{\text{inter}}(\omega) = \frac{e^3 \tau E_0^2}{\pi \hbar^2} \int \Theta(\hbar\omega - \Delta E(\mathbf{k})) \partial_{k_y} \Omega_{vc,\pm}(\mathbf{k}) \, d^2k,$$

$$\Omega_{vc,\pm}(\mathbf{k}) = \sum_{v,c} \frac{\left|\langle v | \partial_{k_x} H \pm i \partial_{k_y} H | c \rangle\right|^2}{2[\Delta E(\mathbf{k})]^2},$$

(19)

where $\pm$ refers to the right and left circularly polarized light, respectively, $\tau$ is the momentum relaxation time, and $E_0$ is the field amplitude of the light.

To calculate $\Omega_{vc,\pm}(\mathbf{k})$, we consider electrons near X valley ($k_x \approx k_X$ and $k_y \approx 0$) and keep first order deviations. Since the Hamiltonian is spin diagonal, we approximate $|c\rangle = |\text{Sn}, \sigma\rangle$, $|v\rangle = |\text{Te}, \sigma\rangle$ and sum the contributions for each spin $\sigma = \pm$. For the orbital part, we may approximate $\langle v | \boldsymbol{\tau} | v \rangle \approx (0, -\alpha_L \hbar k_y / J, -1)$ for the highest-valance band, and $\langle c | \boldsymbol{\tau} | c \rangle = -\langle v | \boldsymbol{\tau} | v \rangle$ for the lowest conduction band. With these projections, $\Delta E(\mathbf{k}) = E_{\text{Te}}(\mathbf{k}) - E_{\text{Sn}}(\mathbf{k}) - 2\bar{\lambda} \sigma \alpha_L \hbar k_y / J$, where the last term is nothing but the spin Rashba effect. We neglect higher order contributions from the ferroelectric polarization. Lastly, we take a parabolic approximation $E_{\text{Te}}(\mathbf{k}) \approx -\hbar^2 \mathbf{k}^2 / 2m_{\text{Te}}$ and $E_{\text{Sn}}(\mathbf{k}) \approx E_{\text{gap}} + \hbar^2 \mathbf{k}^2 / 2m_{\text{Sn}}$ thus $\Delta E(\mathbf{k}) \approx E_{\text{gap}} + \hbar^2 \mathbf{k}^2 / 2m_{vc}$, where $m_{vc}^{-1} = m_{\text{Sn}}^{-1} + m_{\text{Te}}^{-1}$. After some algebra,

$$D_y^{\text{inter}}(\omega) = -d \frac{4 m_{vc} a^2}{\pi \hbar^3 \omega} \left[ t_\sigma^2 - t_\pi^2 + a'(t_\sigma t_{\pi,\Delta} + t_\pi t_{\sigma,\Delta}) \right] \left(1 - \frac{E_{\text{gap}}}{\hbar\omega}\right) \propto \alpha_L \left(1 - \frac{E_{\text{gap}}}{\hbar\omega}\right), \quad (20)$$

$$J_{y,\pm} = \mp d \frac{16 m_{vc} e^3 \tau E_0^2 a^2}{2 \hbar^5 \omega} \left[ t_\sigma^2 - t_\pi^2 + a'(t_\sigma t_{\pi,\Delta} + t_\pi t_{\sigma,\Delta}) \right] \left(1 - \frac{E_{\text{gap}}}{\hbar\omega}\right) \propto \alpha_L \left(1 - \frac{E_{\text{gap}}}{\hbar\omega}\right). \quad (21)$$

Since the ferroelectric polarization $P$ is proportional to $d$, we obtain the following rules.

$$J_{y,+}(P) = -J_{y,-}(P),$$
$$J_{y,\pm}(P) = -J_{y,\pm}(-P).$$

(22)

which is consistent with the charge photocurrent of Fig. 5g in the main text.

Next, we calculate the spin photogalvanic effect. The spin current is written as

$$J_{y,s,\pm} = \frac{e^3 \tau E_0^2}{\pi \hbar^2} \int \Theta(\hbar\omega - \Delta E(\mathbf{k})) \partial_{k_y} \Omega_{vc,s,\pm}(\mathbf{k}) \, d^2k,$$

$$\Omega_{vc,s,\pm}(\mathbf{k}) = \sum_{v,c} \frac{\langle v|\{\sigma_z, \partial_{k_x} H \pm i\partial_{k_y} H\}|c\rangle \langle c|\partial_{k_x} H \mp i\partial_{k_y} H|v\rangle}{4[\Delta E(\mathbf{k})]^2} \quad (23)$$

$$\approx \sum_{v,c} \sigma \frac{|\langle \text{Te}, \sigma|\partial_{k_x} H \pm i\partial_{k_y} H|\text{Sn}, \sigma\rangle|^2}{2[\Delta E(\mathbf{k})]^2}.$$

With the same approximation, we obtain the following spin current.

$$J_{y,s,\pm} = \alpha_L \frac{8 m_{vc} \bar{\lambda} e^3 \tau E_0^2 a^2}{E_{\text{gap}}^2 J \hbar^5 \omega^2} (t_\sigma^2 + t_\pi^2)$$

$$\times \left[ \frac{3a^2 (\hbar\omega - E_{\text{gap}})^2 E_{\text{gap}} m_{vc}}{\hbar^2} \cos^2 \frac{k_X a'}{2} + 2(\hbar^2 \omega^2 - E_{\text{gap}}^2) \sin^2 \frac{k_X a'}{2} \right], \quad (24)$$

which implies

$$J_{y,s,+}(P) = J_{y,s,-}(P),$$
$$J_{y,s,\pm}(P) = -J_{y,s,\pm}(-P), \quad (25)$$

and fully consistent with the spin part of Fig. 5g in the main text.

## Supplementary References

1    Xu, S.-Y. et al., Electrically switchable Berry curvature dipole in the monolayer topological insulator WTe2. *Nat. Phys.* **14**, 900-906 (2018).

2    Sipe, J. E. & Shkrebtii, A. I. Second-order optical response in semiconductors. *Phys. Rev. B* **61**, 5337-5352 (2000).